\documentclass[fleqn,twoside]{article}
\usepackage{espcrc2}

\usepackage{graphicx}
\usepackage[figuresright]{rotating}

\newcommand{\AmS}{{\protect\the\textfont2
  A\kern-.1667em\lower.5ex\hbox{M}\kern-.125emS}}
\def\eko{\tau^- \to e^- K_S^0}
\def\muko{\tau^- \to \mu^- K_S^0}
\def\ekoko{\tau^- \to e^- K_S^0 K_S^0}
\def\mukoko{\tau^- \to \mu^- K_S^0 K_S^0}
\def\lko{\tau^- \to \ell^- K_S^0}
\def\lkoko{\tau^- \to  \ell^- K_S^0 K_S^0}
\def\lkk{\tau^- \to  \ell^- K^+ K^-}

\def\feko{e^- K_S^0}
\def\fmko{\mu^- K_S^0}
\def\fekoko{e^- K_S^0 K_S^0}
\def\fmkoko{\mu^- K_S^0 K_S^0}

\def\flkk{\ell^- K^+ K^-}

\def\BR{{\cal B}}

\newcommand{\prdd}[3]{Phys. Rev. D {\bf #1}, #3 (#2)}
\newcommand{\epjc}[3]{Eur. Phys. J. C {\bf #1}, #3 (#2)}
\newcommand{\prll}[3]{Phys. Rev. Lett. {\bf #1}, #3 (#2)}
\newcommand{\cpc}[3]{Comput. Phys. Commun. {\bf #1}, #3 (#2)}
\newcommand{\mpll}[3]{Mod. Phys. Lett. A {\bf #1}, #3 (#2)}
\newcommand{\nimps}[3]{Nucl. Instrum. Methods Phys. Res., Sect. A {\bf #1}, #3 (#2)}
\def\etal{{\em et~al.}}
\def\ibid{{\em ibid.}}

\title{Search for neutrinoless $\tau$ decays involving the $K_S^0$ meson}

\author{K.K. Gan\address{The Ohio State University,\\
        Columbus, OH 43210,
        U.S.A.}
        \thanks{Representing the CLEO Collaboration.}}
       
\begin{document}

\begin{abstract}
We have searched for lepton flavor violating decays of the $\tau$ lepton with at least
one $K_S^0$ meson in the final state. The data used in the search were collected with
the CLEO II and II.V detectors at the Cornell Electron Storage Ring (CESR) and correspond
to an integrated luminosity of 13.9 $fb^{-1}$ at the $\Upsilon(4S)$ resonance. No evidence
for signals were found, therefore we have set 90\% confidence level (C.L.) upper limits on
the branching fractions $\BR(\eko) <\ 9.5 \times 10^{-7}$, $\BR(\muko) <\ 9.5 \times 10^{-7}$,
$\BR(\ekoko) <\ 2.4 \times 10^{-6}$, and $\BR(\mukoko) <\ 3.5 \times 10^{-6}$. These represent
significantly improved upper limits on the two-body decays and first upper limits on the
three-body decays.
\vspace{1pc}
\end{abstract}

\maketitle

\section{INTRODUCTION}

In physics, all fundamental conservation laws are expected to have associated symmetries.
Lepton flavor conservation, however, is an experimentally observed phenomena with no
associated symmetry in the Standard Model. Lepton flavor violation (LFV) is expected in
many extensions of the Standard Model such as lepto-quark, supersymmetry, superstring,
left-right symmetric models, and models that include heavy neutral leptons
~\cite{Gonz92,Kim97}. Experimentally, both Super Kamiokande~\cite{Fukuda98} and
SNO~\cite{SNO01} observe neutrino oscillation, which may imply LFV in the neutrino
sector, therefore LFV is expected to occur in charged lepton decay at some branching
fraction, albeit very small. The $\tau$ lepton provides a clean laboratory for such searches.
Ilakovac~\cite{Ilakovac95} has calculated upper limits on branching fractions for many
neutrinoless LFV modes within a model involving heavy Dirac neutrinos. The branching
fractions depend on the heavy neutrino masses and mixings. For the decays $\lko$~\cite{note1},
the branching fractions are of ${\cal O}(10^{-16})$, where $\ell$ can be $e$ or $\mu$. For the
decays $\lkoko$ or $\lkk$, the branching fractions are of ${\cal O}(10^{-7})$.
The decays with two kaons in the final state are therefore of particular experimental interest.
Previous published upper limits on the branching fractions for the decays $\lko$ are of
${\cal O}(10^{-4})$~\cite{Hayes82}. There are no previous results for the decays $\lkoko$. In
this paper, we present the result of a search for the decays into one lepton and one or two
$K_S^0$ mesons, with the $K_S^0$ decaying into two charged pions.

The data used in this analysis were collected using the CLEO detector~\cite{CLEOII,CLEOIIV} from
$e^+e^-$ collisions at the Cornell Electron Storage Ring (CESR) at a center-of-mass energy
$\sqrt s \sim 10.6$ GeV. The total integrated luminosity of the data sample is 13.9 $fb^{-1}$
corresponding to the production of $N_{\tau\tau} = 1.27 \times 10^7$ $\tau^+\tau^-$ events.
The CLEO detector is a general purpose spectrometer with excellent charged
particle and shower energy detection.

\section{EVENT SELECTION CRITERIA}
The $\tau^+\tau^-$ candidate events must contain four or six charged tracks with zero net
charge. To reject beam-gas events, the distance of closest approach of each non-$K_S^0$
track to the IP must be within 0.5 cm transverse to the beam and 5 cm along the beam direction.
The polar angle $\theta$ of each track with respect to the beam must satisfy $|\cos\theta| < 0.90$.
Photons are defined as energy clusters in the calorimeter with at least 60 MeV in the barrel
($|\cos\theta| < 0.80$) or 100 MeV in the endcap ($0.80 < |\cos\theta| < 0.95$). We further
require every photon to be separated from the projection of any charged track by at least 30 cm
unless its energy is greater than 300 MeV. In order to diminish QED background such as radiative
Bhabha and $\mu$-pair events with photon conversion, we require each event to have total energy
less than 95\% of the center-of-mass energy. This requirement rejects most of the QED background,
while incurring a small loss in detection efficiency.

We divide each event into two hemispheres (signal and tag), one containing one charged track
and the other containing three or five charged tracks, using the plane perpendicular to the
thrust axis~\cite{Farhi77}. The thrust axis is calculated from both charged tracks and photons.
The invariant mass of the tag hemisphere must be less than the $\tau$ mass,
$M_{\tau}=1.777$~GeV/$c^2$~\cite{PDG00}. The signal hemisphere must contain an electron or a
muon and one or two $K_S^0$ mesons. The electron candidate must have shower energy to momentum
ratio in the range, $0.85 < E/p < 1.10$, and when available, the specific ionization lost
must be consistent with an electron. The muon candidate must penetrate at least three absorbtion
lengths of iron. The $K_S^0$ candidate is reconstructed in the $\pi^+\ \pi^-$ final state with a
detached vertex, and the invariant mass must be within approximately three standard deviations of the
nominal mass, $485 < m_{\pi \pi} < 510$ MeV/$c^2$. In order to diminish radiative Bhabha
and $\mu$-pair events further, neither pion should be consistent with identification as
an electron.

Since there is no neutrino in the signal hemisphere while there is at least one neutrino
undetected in the tag hemisphere, the missing momentum of the event must point toward the
tag hemisphere, $0 < \cos\theta_{tag-missing} < 1.0$. In order to suppress the background from
radiative Bhabha and $\mu$-pair events, the direction of the missing momentum of the event is
required to satisfy $|\cos\theta_{missing}| < 0.90$. For the decay $\eko$,
$\cos\theta_{tag-missing}$ is further restricted to be less than 0.99 to reduce the radiative
backgrounds. This corresponds to the minimum ratio of background to detection efficiency. The
background is estimated from the sidebands in the invariant mass vs. total energy distribution
of the decay candidates.

\begin{figure}[p]
\centering
\includegraphics[width=2.9in]{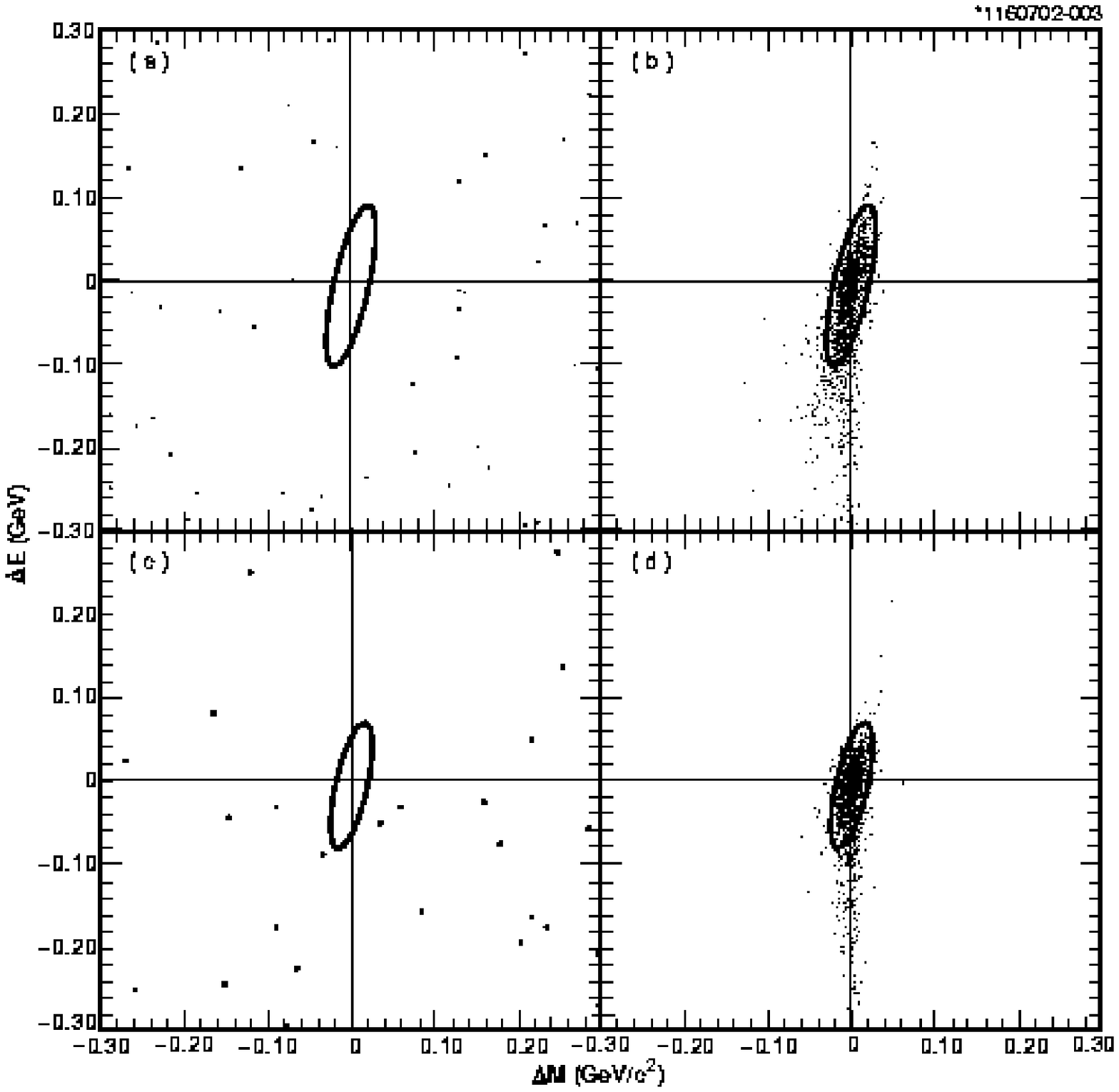}
\vspace{0.5in}
\caption{$\Delta E$ vs. $\Delta M$ distribution of the (a) data and (b) signal Monte Carlo sample for
the decay $\eko$; (c) and (d) show the corresponding distributions for $\muko$. The normalization of
the signal Monte Carlo sample is arbitrary. The ellipses indicate the signal region (see text).}
\label{fig:sbko}
\end{figure}

\begin{figure}[p]
\centering
\includegraphics[width=2.9in]{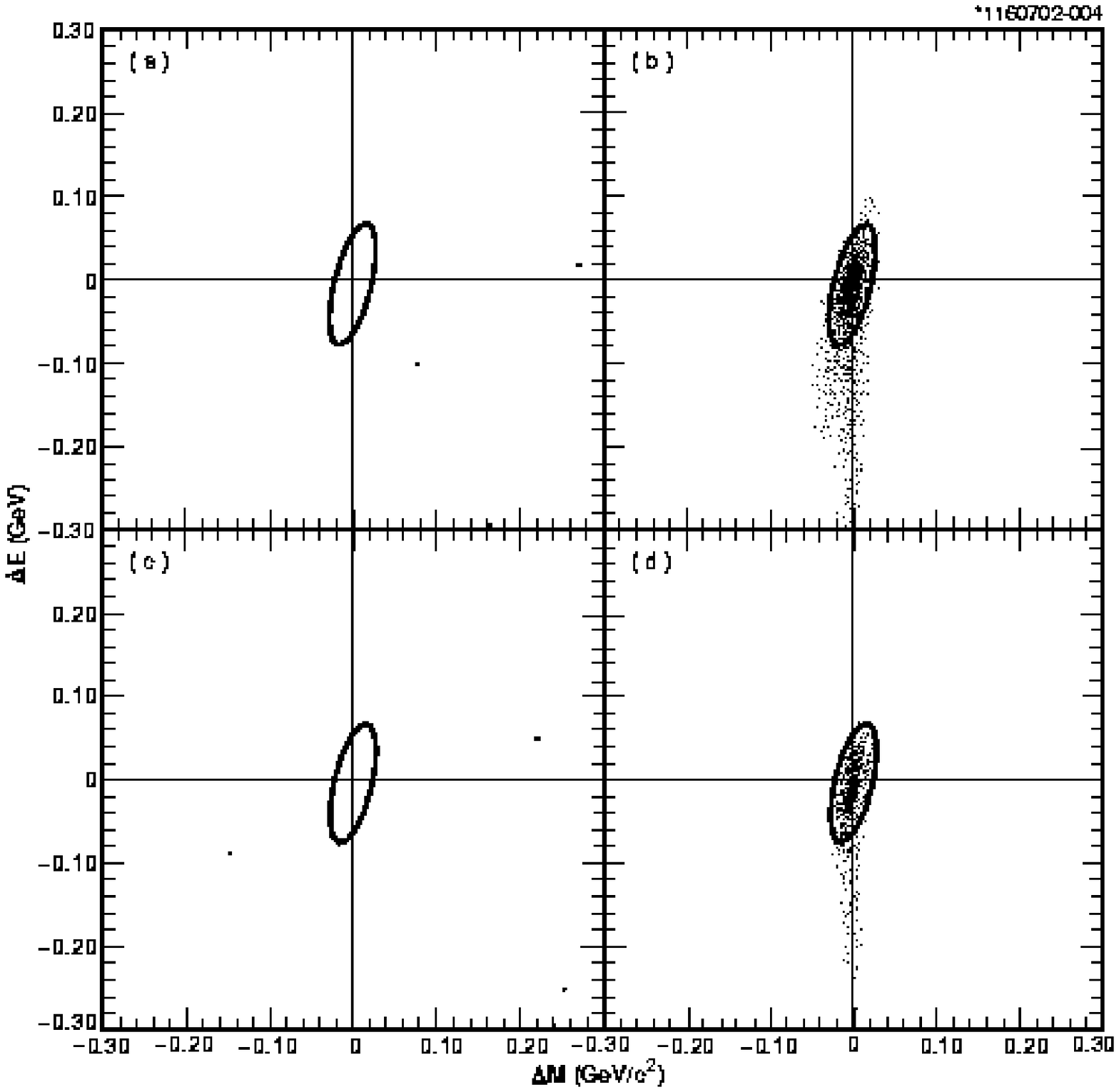}
\vspace{0.5in}
\caption{$\Delta E$ vs. $\Delta M$ distribution of the (a) data and (b) signal Monte Carlo sample for
the decay $\ekoko$; (c) and (d) show the corresponding distributions for $\mukoko$. The normalization of
the signal Monte Carlo sample is arbitrary. The ellipses indicate the signal region (see text).}
\label{fig:sbkoko}
\end{figure}

To search for decay candidates, we select $\tau$ candidates with invariant mass and total energy
consistent with the expectations.  The following kinematic variables are used to select the
candidate events:
\begin{eqnarray*}
\Delta E & = & E - E_{beam} \\
\Delta M & = & M - M_{\tau}\ \ ,
\end{eqnarray*}
where $E_{beam}$ is the beam energy, and $E$ and $M$ are the reconstructed $\tau$ energy and
mass. The $\Delta E$ vs. $\Delta M$ distributions of the decay candidates in the data and
signal Monte Carlo samples are shown in Figs.~\ref{fig:sbko} and \ref{fig:sbkoko}.

In the Monte Carlo simulation, one $\tau$ lepton decays according to two- or three-body phase space
for the mode of interest, and the other $\tau$ lepton decays generically according to the
$KORALB/TAUOLA$ $\tau$ event generator~\cite{taugen85-91}.
The phase space model is appropriate for an unpolarized tau. If the Lorentz structure of the
neutrinoless decay is V-A (Standard Model-like), correlations between the spins of the two
$\tau$'s in the event will lead to slightly higher detection efficiency than phase space,
while V+A decays will lead to lower detection efficiency.
The detector response is simulated using the $GEANT$ program~\cite{Brun-GEANT}. The estimated detection
efficiencies ($\epsilon$)~\cite{eff} are summarized in Table~\ref{tab:results}.

\begin{table}[th]
\begin{centering}
\caption[]{Summary of detection efficiency (with statistical error), 90\% C.L. upper limits on the branching
fraction with and without including systematic errors.}
\label{tab:results}
\vspace{0.1in}
\begin{tabular}{lccc}
\hline
Mode       & $\epsilon$ (\%)    & $\BR(10^{-7})$         & $\BR(10^{-7})$ \\
           &                    &  (stat.)               &   \\
\hline
$\feko$    & $18.8 \pm 0.4$     &  8.8                   &  9.5 \\
$\fmko$    & $19.0 \pm 0.4$     &  8.7                   &  9.5 \\
$\fekoko$  & $12.0 \pm 0.1$     &  21                    &  24  \\
$\fmkoko$  & $ 8.0 \pm 0.1$     &  31                    &  35  \\
\hline
\end{tabular}
\end{centering}
\end{table}

\section{RESULTS}
The upper limit on the branching fraction is related to the upper limit $\lambda$ on the number of
signal events by
\begin{eqnarray}
\nonumber
{\cal B} =
\frac { \lambda}{ 2 \epsilon N_{\tau\tau} {\cal B}_1 ({\cal B}_{K_S^0 \to \pi^+\pi^-})^n}\ \ ,
\end{eqnarray}
where ${\cal B}_1 = (84.71 \pm 0.13)\% $ is the inclusive 1-prong branching fraction \cite{PDG00},
${\cal B}_{K_S^0 \to \pi^+\pi^-} = (68.61 \pm 0.28)\% $ is the branching fraction for $K_S^0$ to
decay to two charged pions, and $n$ is the number of $K_S^0$ mesons in the final state. No candidate
decays are observed, so we take $\lambda$ as 2.44 events in each mode at 90\% confidence level,
according to the frequentist method~\cite{Feldman98}. The upper limits on the branching fractions
with and without systematic errors are shown in Table~\ref{tab:results}.

The systematic errors include the uncertainties in the $\tau^+\tau^-$
cross section (1\%), luminosity (1\%), track reconstruction efficiency (1\% per charged track),
$K_S^0$ detection efficiency (2\% per $K_S^0$), lepton identification (1.5\% for electron and 4\%
for muon), and the statistical uncertainties in the detection efficiencies due to limited Monte
Carlo samples (1-2\%).

Black \etal~\cite{Black02} have analyzed the constraints on the new physics scale for dimension-six
fermionic effective operators involving tau-mu mixing, motivated by the observed $\nu_\mu - \nu_\tau$
oscillation.  The most stringent lower limits from exotic heavy quarks and tau decays on the physics
scale of the operators involving quarks are $\sim$10 TeV.  The new upper limit on $\BR(\muko)$ presented
in this paper yields a lower limit of 17.3 and 18.2 TeV for the axial vector and pseudoscalar operators,
respectively.

In conclusion, we have searched for $\tau$ decays involving $K_S^0$ mesons that violate lepton
flavor, but find no evidence for a signal. This results in improved upper limits for the
decays $\lko$ and first upper-limits for the decays $\lkoko$. The upper limits for the
$\lkoko$ final states are more stringent than those found previously for $\flkk$~\cite{Bliss98}.

This work was supported in part by the U.S. Department of Energy.


\begin{thebibliography}{9}
\bibitem{Gonz92}
M.C.~Gonzalez-Garcia and J.W.F.~Valle, \mpll{7}{1992}{477}; A {\bf 9}, 2569 (E) (1994).

\bibitem{Kim97}
J.E.~Kim \etal, \prdd{56}{1997}{100}.

\bibitem{Fukuda98}
Super Kamiokande Collaboration, Y.~Fukuda \etal, \prll{81}{1998}{1562}.

\bibitem{SNO01}
SNO Collaboration, Q.R.~Ahmad \etal, \prll{87}{2001}{071301}.

\bibitem{Ilakovac95}
A.~Ilakovac \etal, \prdd{52}{1995}{3993}; \prdd{62}{2000}{036010}.

\bibitem{note1}
Throughout this paper, the charge conjugate state is implied.

\bibitem{Hayes82}
MARK II Collaboration, K.G.~Hayes \etal, \prdd{25}{1982}{2869}.

\bibitem{CLEOII}
Y.~Kubota \etal, \nimps{320}{1992}{66}.

\bibitem{CLEOIIV}
T.~Hill, \nimps{418}{1998}{32}.

\bibitem{Farhi77}
E.~Farhi, \prll{39}{1977}{1587}.

\bibitem{PDG00}
Particle Data Group, J.~Bartels \etal, \epjc{15}{2000}{1}.

\bibitem{taugen85-91}
S.~Jadach and Z.~Was, \cpc{36}{1985}{191}; {\bf 64}, 267 (1991); S.~Jadach, J.H.~Kuhn,
and Z.~Was, \ibid~{\bf 64}, 275 (1991).

\bibitem{Brun-GEANT}
R.~Brun \etal, CERN Report No. CERN-DD/EE/84-1, 1987 (unpublished).

\bibitem{eff}
The detection efficiencies for $\mukoko$ depends on the mass of the $K_S^0$-pair produced.
The detection efficiency is approximately constant up to $M_{K_S^0K_S^0} \sim 1.25$ GeV/$c^2$,
falling to zero near the kinematic limit, $M_{K_S^0K_S^0} = M_\tau$. The kinematic limit
corresponds to the muon being produced at rest in the center-of-mass frame of the $\tau$-lepton.
In the laboratory frame, the muon has low momentum, hence would not be able to penetrate enough material
to be classified as a muon.

\bibitem{Feldman98}
G.J.~Feldman and R.D.~Cousins, \prdd{57}{1998}{3873}.

\bibitem{Black02}
D.~Black, T.~Han, H.J.~He and M.~Sher, \prdd{66}{2002}{053002}.

\bibitem{Bliss98}
CLEO Collaboration, D.W.~Bliss \etal, \prdd{57}{1998}{5903}.

\end{thebibliography}
\end{document}